# Magnetism in $AV_3Sb_5$ (A = Cs, Rb, K): Complex Landscape of the Dynamical Magnetic Textures

Debjani Karmakar[1,2,#,*], Manuel Pereiro[1,#], Md. Nur Hasan[3], Ritadip Bharati[4], Johan Hellsvik[5], Anna Delin[6,7], Samir Kumar Pal[3], Anders Bergman[1], Shivalika Sharma[8], Igor Di Marco[1,8,9], Patrik Thunström[1], Peter M. Oppeneer[1] and Olle Eriksson[1,*]

[1] *Department of Physics and Astronomy,*
*Uppsala University, Box 516, SE-751 20, Uppsala, Sweden*

[2] *Technical Physics Division*
*Bhabha Atomic Research Centre, Mumbai 400085, India*

[3] *Department of Chemical and Biological Sciences,*
*S. N. Bose National Centre for Basic Sciences,*
*Block JD, Sector-III, SaltLake, Kolkata 700 106, India*

[4] *School of Physical Sciences*
*National Institute of Science Education and Research*
*HBNI, Jatni - 752050, Odisha, India*

[5] *PDC Center for High Performance Computing,*
*KTH Royal Institute of Technology, SE-100 44 Stockholm, Sweden*

[6] *Department of Applied Physics*
*KTH Royal Institute of Technology, SE-106 91 Stockholm, Sweden*

[7] *Swedish e-Science Research Center (SeRC),*
*KTH Royal Institute of Technology, SE-10044 Stockholm, Sweden*

[8] *Asia Pacific Center for Theoretical Physics,*
*Pohang, 37673, Republic of Korea*

[9] *Department of Physics, Pohang University of Science and Technology,*
*Pohang, 37673, Republic of Korea*

[#] D.K. and M.P. equally contributed

[*]Corresponding Authors:
    Debjani Karmakar
    Email: debjani.karmakar@physics.uu.se

    Olle Eriksson
    Email: olle.eriksson@physics.uu.se

**Abstract**

We have investigated the dynamical magnetic properties of the V-based kagome stibnite compounds by combining the *ab-initio* calculated magnetic parameters of a spin Hamiltonian like inter-site exchange parameters, magnetocrystalline anisotropy and site projected magnetic moments, with full-fledged simulations of atomistic spin-dynamics. Our calculations reveal that in addition to a ferromagnetic order along the [001] direction, the system hosts a complex landscape of magnetic configurations comprised of commensurate and incommensurate spin-spirals along the [010] direction. The presence of such chiral magnetic textures may be the key to solve the mystery about the origin of the experimentally observed inherent breaking of the $C_6$ rotational symmetry- and the time-reversal symmetry.

For the V-based kagome stibnites, $AV_3Sb_5$ (A=Cs, Rb, K), a complete picture of the origin of the experimentally observed spontaneously broken time-reversal as well as $C_6$ rotational symmetries and their respective interplay with the underlying magnetic structure is missing. This lacuna has proliferated into wide variations of inconclusive explanations towards the experimental results like muon spin-rotation (μSR) or anomalous Hall Effect (AHE) [1,2]. In the zero-field μSR spectra, the depolarization function of single crystals of these materials deviate from the standard Gaussian Kubo-Toyabe behavior, for which the possible reasons may be (a) an inhomogeneous distribution of the nuclear moments, (b) the presence of an electric field gradient due to a chiral charge-order (CO) or (c) a contribution of purely electronic origin emanating from the chiral distribution of the spin moments [1]. Below the CO transition and at high-field, the significant increase of the rate of spin-relaxation of muons and the enhanced spread of the internal fields are suggested to be the experimental signatures of a broken time-reversal symmetry enrooted to their electronic attributes [1]. In the same vein, the experimentally observed giant AHE for this series exhibits an anomalous Hall ratio, which is an order of magnitude higher than that of intrinsic magnetic systems like bcc Fe [3]. The nature of AHE is proposed to be of extrinsic type related to the spin-fluctuations across the triangular V-clusters leading to an enhanced skew scattering [2]. Currently proposed microscopic mechanisms behind the nature of the μSR spectra, the giant AHE and their interconnection with the broken $C_6$ rotational- and time-reversal symmetries are far from conclusive, suggesting a wide varieties of reasons, *viz*., the presence of orbital current, electron nematicity or a 2 × 2 charge modulation producing chiral charge density waves [1-3]. The effects of spin-density wave, complex spin-texture and dynamical magnetic ground states on the CO and the broken symmetries therein are yet to get analyzed.

In recent literatures, there are controversies regarding the presence of broken time-reversal symmetry in the first member of this kagome superconducting series, *viz*., $CsV_3Sb_5$. Saykin et al. [4] have used high resolution polar Kerr effect measurements to demonstrate an absence of any observable Kerr effect in $CsV_3Sb_5$, which poses a question-mark against the presence of spontaneous breaking of time-reversal symmetry in this compound. On the other hand, albeit being centrosymmetric, this system reveals a chiral transport behavior via second-harmonic generation under an in-plane applied magnetic field [5] implying an inherently broken mirror symmetry. Such occurrence of electronic magnetochiral anisotropy is prominent below 35K and thereby indicates interplay of time-reversal symmetry breaking and electronic chirality in $CsV_3Sb_5$ below this transition temperature.

The two-dimensional (2D) kagome lattice with nearest neighbor (NN) antiferromagnetic (AFM) exchange interactions ($J$) is a well-known archetype of inherently frustrated systems, where, it is impossible to optimize all exchange interactions. Here, after including diverse NN interactions, panoply of degenerate magnetic ground states may arise, where the out-of-plane AFM or ferrimagnetic order competes with the co-planar $120^0$ structures with uniform or staggered chirality [6, 7]. Depending on the temperature ($T$), their magnetic ground states encompass three different regimes, *viz.*, i) a spin-diffusive paramagnetic regime ($T > |J|$), ii) an intermediate cooperative regime leading to a spin-liquid arrangement ($0.001|J| < T < |J|$) and iii) an ultralow temperature coplanar regime ($T < 0.001|J|$). The first crossover at $T \sim J$ may have a finite local spin-correlation without any long-range order (LRO) and for the second one at $T \sim 10^{-3}J$, the entropic selection favors a coplanar order with anisotropic dynamics [8-11]. For the coplanar configuration, thermal or quantum fluctuations resolve the ground-state degeneracy after stabilizing a LRO in the *order by disorder* process [11], the most common of which is the $120^0$ spin-texture. With increasing temperature, such chiral LRO is intervened after generation of weathervane defects leading to the rotational staggering of spins around the local spin-axis [6].

The series AV$_3$Sb$_5$, as suggested from their complex magnetic spectral functions [12], is an itinerant-electron system with a correlated electronic structure. The low-energy magnetic excitations of such systems can be evaluated via a multi-scale approach upon satisfaction of three conditions, *viz.*, 1) the magnetic configurational energy dependence can be mapped onto a generalized Heisenberg model-Hamiltonian, $H = -\sum_{i \neq j} e_i^\alpha \hat{J}_{ij}^{\alpha\beta} e_j^\beta$, $\alpha, \beta = x, y, z$, with the unit vector $e_i$ designating the local spin direction at the *i*-th site, 2) the exchange tensor $\hat{J}_{ij}^{\alpha\beta}$ can be calculated from the first-principles and 3) the ground state many-electron systems relies on the adiabatic approximations, with decoupled faster process of inter-site electronic hopping from the slower moving magnetic excitations or "frozen" magnons [13]. In our prior study [12], the fully relativistic exchange tensor was reported using the Lichtenstein-Katnelson-Antropov-Gubanov (LKAG) formalism coupled to the dynamical mean field theory (DMFT) [14-20]. The tensor possesses terms like the symmetric-isotropic Heisenberg($J_{ij}$), anti-symmetric and anisotropic Dzyaloshinskii-Moriya (DM) and symmetric anisotropic Heisenberg interactions ($\Gamma$). The DM and $\Gamma$ interactions are calculated by using the relations like; $D_{ij}^z = \frac{1}{2}(\hat{J}_{ij}^{xy} - \hat{J}_{ij}^{yx})$, $\Gamma_{ij}^z = \frac{1}{2}(\hat{J}_{ij}^{xy} + \hat{J}_{ij}^{yx})$ and analogous ones for the *x*- and *y*-components. Fig. 1 presents the plots of the inter-site exchange interactions for various shells of NN of V-based stibnites extracted with an FM interlayer interaction. The negative values of $J_{ij}$ correspond to an AFM coupling. A closer scrutiny of this figure reveals that the magnetic interactions in this system are composed of various competing mechanisms. For the first NN shell, the $J_{ij}$ values are roughly two orders of magnitude larger than the $D_{ij}$ and $\Gamma_{ij}$ values. However, from the next NN-shell, the anisotropic interactions $D_{ij}$ and $\Gamma_{ij}$ becomes comparable to the $J_{ij}$'s. Such a competing environment of complex magnetic interactions can be a fertile ground for obtaining chiral magnetic ground-states. It may be mentioned in passing that for systems like metallic and Mott insulating kagome systems, chiral magnetism may appear from very simple correlated Hamiltoniian [21-22]. Moreover, at the same NN distances, the flipping sign of the $D_{ij}$'s implies a signature of the chiral magnetic textures as per the Moriya rule.

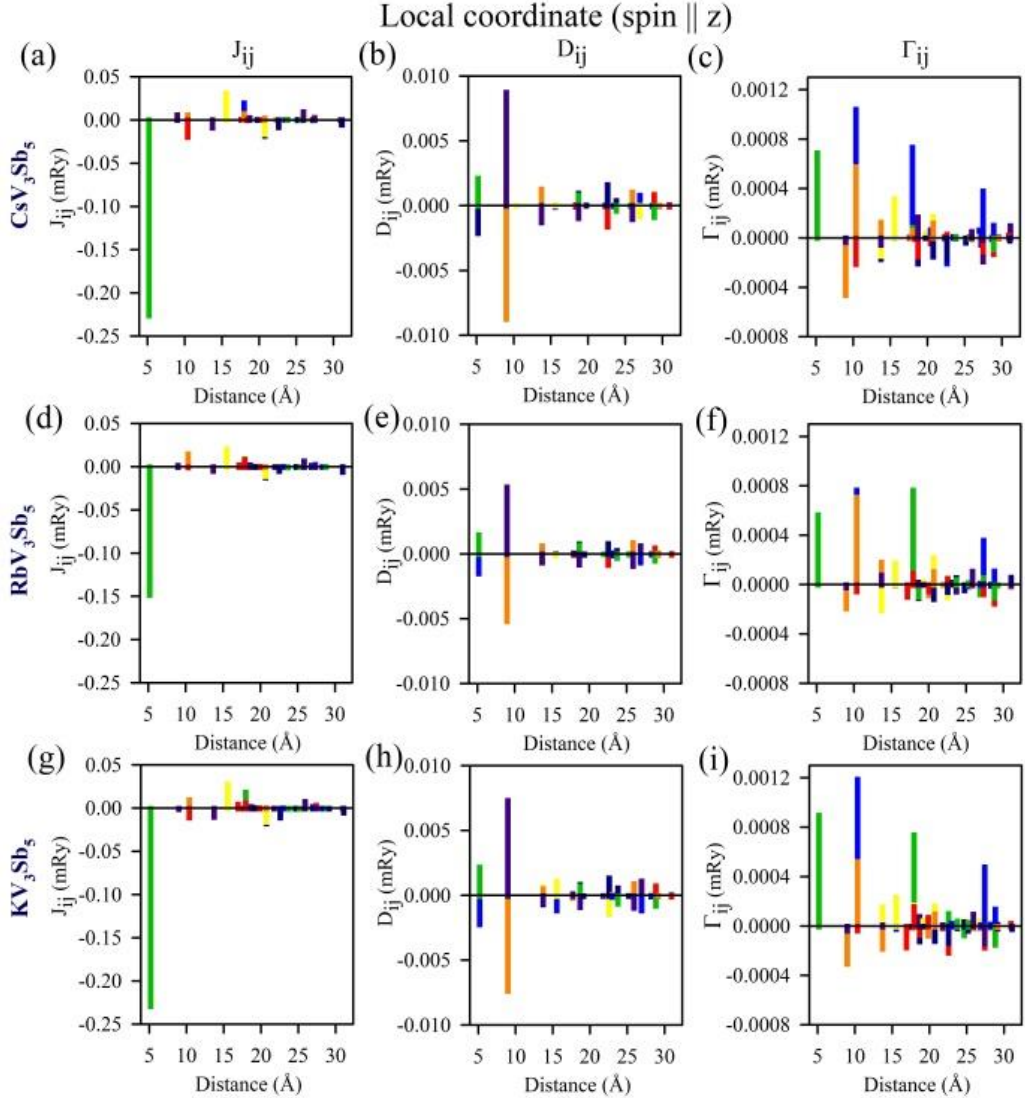

FIG. 1: The inter-site exchange parameters as a function of NN distances in local coordinate system: $J_{ij}$, $D_{ij}$ and $\Gamma_{ij}$ for (a)-(c) $CsV_3Sb_5$; (d)-(f) $RbV_3Sb_5$ and (g)-(k) $KV_3Sb_5$. Note that, for some NN distances, we have different strength of the interactions in different directions, implying anisotropy in the directional distribution of the interactions. The difference in colour at the bar plot signifies the difference of the strength of various magnetic exchange interactions at a particular value of NN distance.

To obtain inversion asymmetric chiral environments with significant $D_{ij}$ and $\Gamma_{ij}$, systems were artificially designed using metallic multilayers like Mn/Cr/W[110] or Pd-Fe/Ir[111] involving components of high spin-orbit coupling (SOC) [23-26]. Recently, such spiral magnetic textures have gained attention by virtue of their additional capability to explore the topological spin-transport via skyrmions [25,26] for spin-torque devices [27-29]. Here, the limitations introduced by the instabilities like Walker breakdown on the mobility of the domain walls can be escaped by reaching the super-magnonic regime [28-32]. A sustained search for such chiral magnetic systems constitutes an active area of research. In this letter, we demonstrate that the V-based stibnite kagome system, with the presence of local moments and a balanced competition between the inter-site exchanges and SOC, constitutes a breeding ground for long-period, inhomogeneous chiral magnetic superstructures.

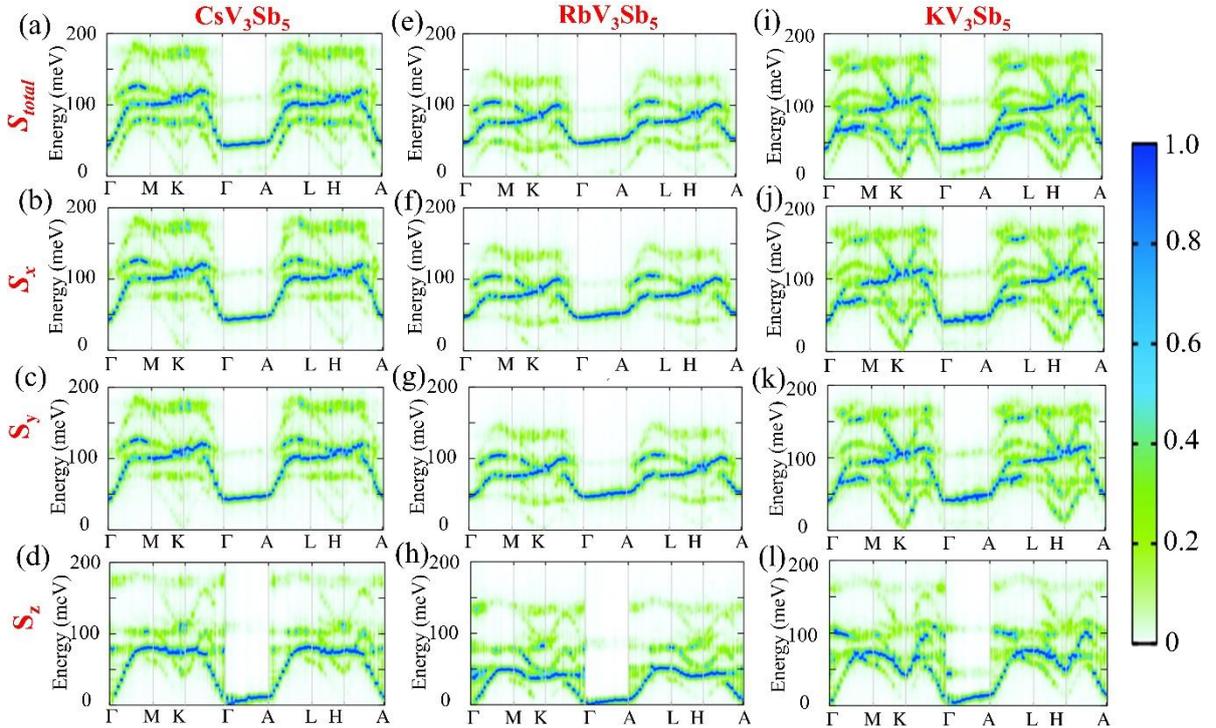

FIG. 2: The dynamical structure factors $S(q,\omega)$ plotted along the high-symmetry paths, (a), (e) and (i) the resultant total, (b), (f) and (j) $S_x$-projected, (c), (g) and (l) $S_y$-projected, (d), (h) and (l) $S_z$-projected $S(q,\omega)$ for $CsV_3Sb_5$, $RbV_3Sb_5$ and $KV_3Sb_5$, respectively. The dynamical spin-correlations shown in the figure are those around the magnetic excitation energy minimum at the K-point.

The ground-state magnetic properties of complex magnetic systems are derived after constructing a classical atomistic spin-model, where the bilinear effective Hamiltonian can be written as: $\mathcal{H}_{mod}^{i,j} = -J_{ij}\boldsymbol{e}_i \cdot \boldsymbol{e}_j - \boldsymbol{D}_{ij}\boldsymbol{e}_i \times \boldsymbol{e}_j - \boldsymbol{e}_i\Gamma_{ij}\boldsymbol{e}_j - \kappa\sum_{k=i,j}(\boldsymbol{e}_k \cdot \boldsymbol{e}_k^r)^2$ [27,33,34]. Here $\boldsymbol{e}_i$ and $\boldsymbol{e}_j$ are normalized atomic spin moments at the atomic sites $i$ and $j$ and $\boldsymbol{e}_k^r$ is the easy axis, along the arbitrary unit vector *r*. The four different terms of this Hamiltonian represent the energies corresponding to the isotropic Heisenberg interaction (J), the DM interaction (DM), the symmetric anisotropic exchange interaction (Γ) and the magnetocrystalline anisotropy (K) [24].

In a periodic lattice, the most general solution of this model generates a homogeneous spin-spiral, where the angle between the resultant spin-moments of two adjacent neighbours are constant throughout the spiral [35]. The atomic-site (*i*) spin-moments of a such spin-spiral can be mapped onto a unit circle with $1 \leq i \leq N$, with N being the number of atomic spins in a spiral, the pitch of which can be defined as $\lambda = Na$, with *a* being the lattice constant. The chirality of the spiral is dependent on the clockwise (negative) or anti-clockwise (positive) directionality of the mapping. Naturally, the DM interaction has a significant role in determining the chirality of the spiral [23-26].

The presence of SOC and a large number of magnetic atoms render the first-principles treatment of a long-pitched spin-spiral computationally taxing [7,29]. Therefore, for chiral magnetic systems [24,36], a continuum micro-magnetic model is defined, where the energy of the spin-spiral is written in terms of the local magnetization vector *m* as $E[m] = \frac{1}{\lambda}\int_0^\lambda dx \left[\frac{A}{4\pi^2}(\dot{m})^2 + \frac{D}{2\pi}.(m \times \dot{m}) + m^T \kappa m\right]$. Here *A*, *D* and $\kappa$ are the spin-stiffness constant, the effective DM vector and the anisotropy tensor respectively, while $\dot{m}$ is the gradient of the magnetization. For a generalized spin-spiral with a wave vector *q*, this magnetic moment *m* at a given atomic position $R_{n\alpha}$ can be written in the tensor format as $m_{n\alpha} = m_\alpha \begin{pmatrix} sin\theta_\alpha cos(\varphi_{n\alpha} + \gamma_\alpha) \\ sin\theta_\alpha sin(\varphi_{n\alpha} + \gamma_\alpha) \\ cos\theta_\alpha \end{pmatrix}$ [5,24]. Here $\theta_\alpha$ and $\gamma_\alpha$ are the cone and the phase angles respectively. The azimuthal angle corresponding to the local magnetic moment can be calculated as, $\varphi_{n\alpha} = q.R_{n\alpha}$. Such spiral magnetic ground states can be either commensurate or incommensurate with the underlying lattice, the solution of which can be obtained by following the Petit's method [37], using a series of effective coordinate transformations of the local spins into a ferromagnetic (FM) order [37,38].

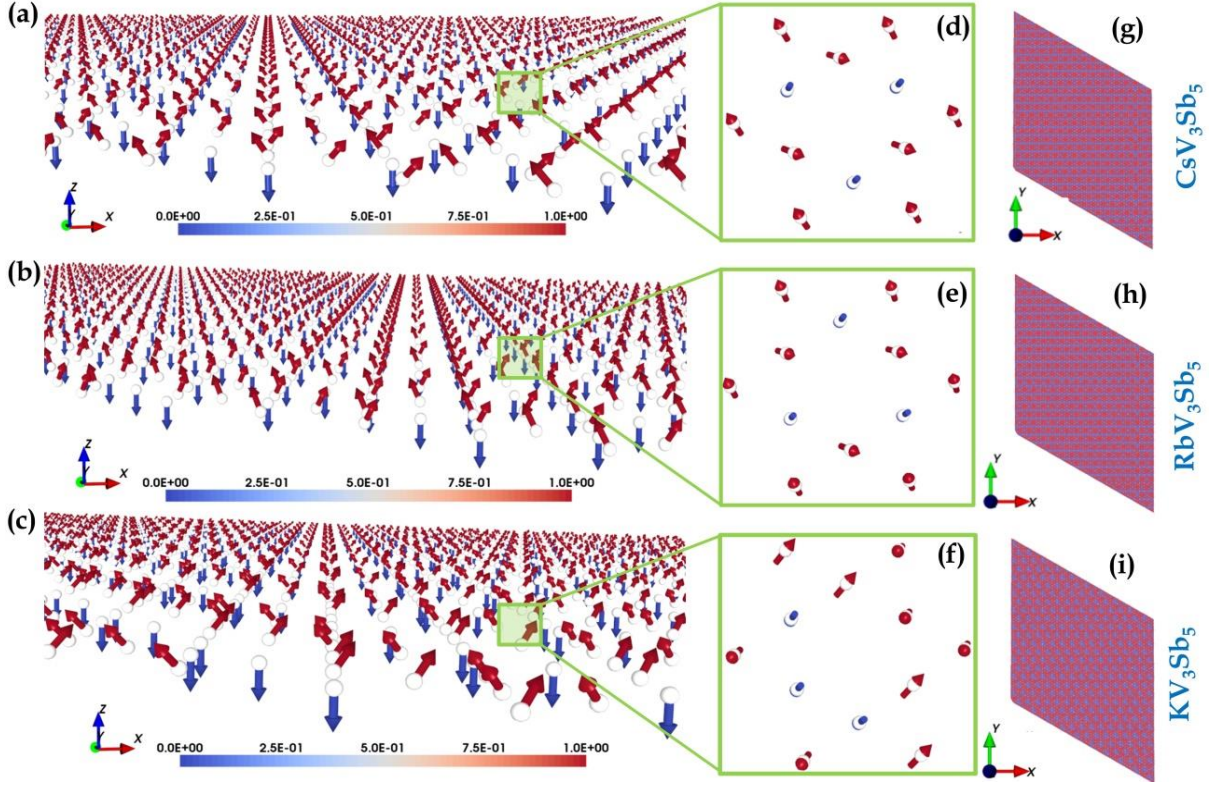

FIG. 3: (a)-(c) The 3D view of the dynamical ground-state spin-textures obtained from the $J + K + \Gamma + DM$ Hamiltonian; (d)-(f) spin-configurations on the star-of-David of the kagome lattice; (g)-(i) the corresponding magnetization densities for CVS, RVS and KVS respectively. The colour bars associated with the spin-textures indicate the projection of the magnetization along $z$-direction. The magnetic textures correspond to the minimum of magnetic excitation energy at the zone-centre ($\Gamma$-point).

For the present kagome-series, the atomistic descriptions of the spin-moments can be used to construct the spin-model because of its sufficiently localized electronic densities around the atomic sites and the invariance of the resultant atomic spin-moments with respect to their orientations. The chiral ground states can be obtained by dynamically solving the atomistic model, where, at every magnetic site, around the direction of the classical spin, the dynamics of the small fluctuations of the spins are calculated, following the implementations in the UppASD software [34].

The spin-spin correlation functions, and their Fourier transform, alias the dynamical structure factor, also accessible experimentally by inelastic neutron scattering, can be calculated as $S(q,\omega) = \frac{1}{2\pi N}\sum_{i,j} e^{iq(r_i-r_j)} \int_{-\infty}^{\infty} d\tau e^{-i\omega\tau} \langle e_i e_j(\tau) \rangle$, where, $r_i$ is the position vectors of the magnetic atoms and $e_i$ is the local spin-vectors. For the incommensurate spin structures, in addition to $\omega(q)$, the magnon dispersion also constitutes modes like $\omega(q \pm Q)$ and the corresponding correlations $S(q,\omega)$ and $S(q \pm Q,\omega)$ describes rigid rotations on the ordering plane and the canting with respect to the same plane respectively [38], where $Q$ is the ordering wave-vector.

Figure 2(a)-(i) represent the spin-component projected dynamical structure factors $S(q,\omega)$ calculated for the $q$-values along the high-symmetry paths for CsV$_3$Sb$_5$ (CVS), RbV$_3$Sb$_5$ (RVS) and KV$_3$Sb$_5$ (KVS) respectively after considering the full Hamiltonian. The highest values of the spin-correlations are obtained along the adiabatic magnon dispersion lines, which in Fig.2 are represented by the transition in colour from green to blue in the adjacent colour-scale. The kagome system in its unit cell possesses 3 V atoms and in the full non-collinear manner, the three possible magnetic orientations should give rise to nine modes. However, along Γ-A direction, the out-of-plane magnon dispersions are triply degenerate. In addition, the *x* and *y* components of the structure factor show the minimum of the energy dispersion at the K point rather than the Γ point, indicating that their magnetic ground-state collapses into a spin-spiral state. This state is shown in Fig. 3 (a-c), and is analysed in more detail below. The *x* and *y* components of $S(q,\omega)$ are seen to have significant values of the magnon gap at the Γ-point. The *z*-component and thus the total dynamical structure factor are seen to display a much smaller value of the gap. The reason behind this gap is the high value of the magnetocrystalline anisotropy, amounting to 0.29 mRy, 0.22 mRy and 0.28 mRy per V atom for CVS, RVS and KVS respectively, as per the DFT + DMFT + SOC calculations. The derived anisotropy

parameters indicate that all three systems display a uniaxial, out-of-plane easy-axis pattern of spin-orientations.

The spin-textures of the dynamically obtained magnetic ground states after solving the full Hamiltonian with all J, DM, $\Gamma$ and K terms are plotted in Fig. 3(a)-(c). The corresponding spin-arrangements in the star-of-David kagome pattern are plotted in Fig. 3(d)-(f) with the magnetization densities in the Fig. 3(g)-(i). The colour scale of the Fig 3 represents the values of the normalized out-of-plane (*z*) projections of the spins. This figure evidences that the dynamical magnetic ground states of CVS and RVS are similar, manifesting a superposition of three types of underlying magnetic orders, *viz.* a FM arrangement along the [001] direction and two spin-spiral configurations, *viz.* one commensurate and the other incommensurate spin spiral arrangements along [010]. The pitches corresponding to the commensurate and the incommensurate spin-spirals are calculated to be 3*a* and 1.5*a* respectively, with *a* being the lattice parameter. The ground-state spin-texture of KVS, as seen in Figure 3(c), is consisted of additional complexities, *viz.* a superposition of *i*) FM order along [001], *ii*) one commensurate spin-spiral along [010] and *iii*) two more inter-penetrating commensurate spirals along [010]. The pitches of all these spirals are 3*a*. In Fig S1 and S2 of supplementary materials (SM), the calculations of the pitches are explained [39]. In Fig. 4(a)-(c), the ground-state commensurate and incommensurate spin-spiral textures are displayed after dynamically solving the full Hamiltonian. Here, the adjacent non-collinear atomistic spins are oriented at an angle of $120^0$ with respect to each other and thus the magnetic ground-state constitutes a three-dimensional (3D) $120^0$ spin-configuration. In Fig S3 of SM, the 3D $120^0$ spin-configurations are illustrated. To demonstrate the importance of the complete Hamiltonian in achieving the ground-state, we have presented the spin-spiral textures by solving the spin-Hamiltonian with several combinations, *viz.*, only J (Fig 4(d)-(f)), J + K (Fig 4(g)-(i)) and J + $\Gamma$ (Fig 4(j)-(l)). The corresponding dynamical structure factors $S(q, \omega)$ with the detailed spin-textures are plotted in Fig S4 – S9 of SM [39].

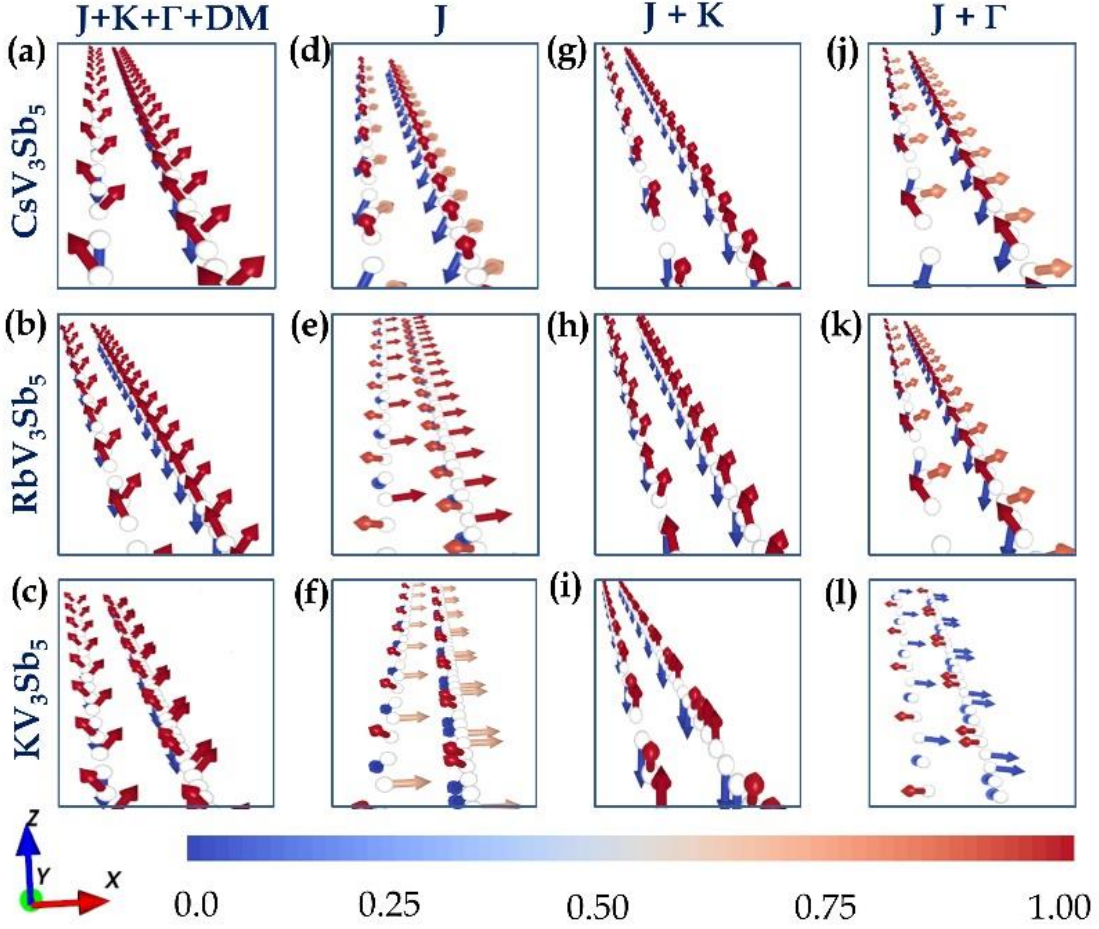

FIG. 4: (a)-(c) The 3D view of the commensurate and incommensurate spirals with J + K + Γ + DM Hamiltonian, (d)-(f) spin-spirals with Hamiltonian having only J, (g)-(i) with J + K, (j)-(l) with J + Γ for CVS, RVS and KVS respectively. For KVS, all the spirals are commensurate. The colour bar indicates the projection of the magnetization along the *z*-direction.

A detailed observation of Fig. 4 evinces that albeit the pitches and the overall types of the spin-spirals remains the same for all three systems for all combinations, the 3D $120^0$ spin-configurations are stabilized only with the complete Hamiltonian. All of these spin-dynamical ground states are simulated at an ultralow temperature of 0.001K. With increasing temperature, we have investigated the evolution of the spin-textures in Fig 5, where the temperature-induced increase of configurational entropy leads to a prominent presence of weathervane defects [39].

However, the spin-spiral textures remain intact up to 5K, which is higher than the superconducting critical temperature, $T_c$ of this series.

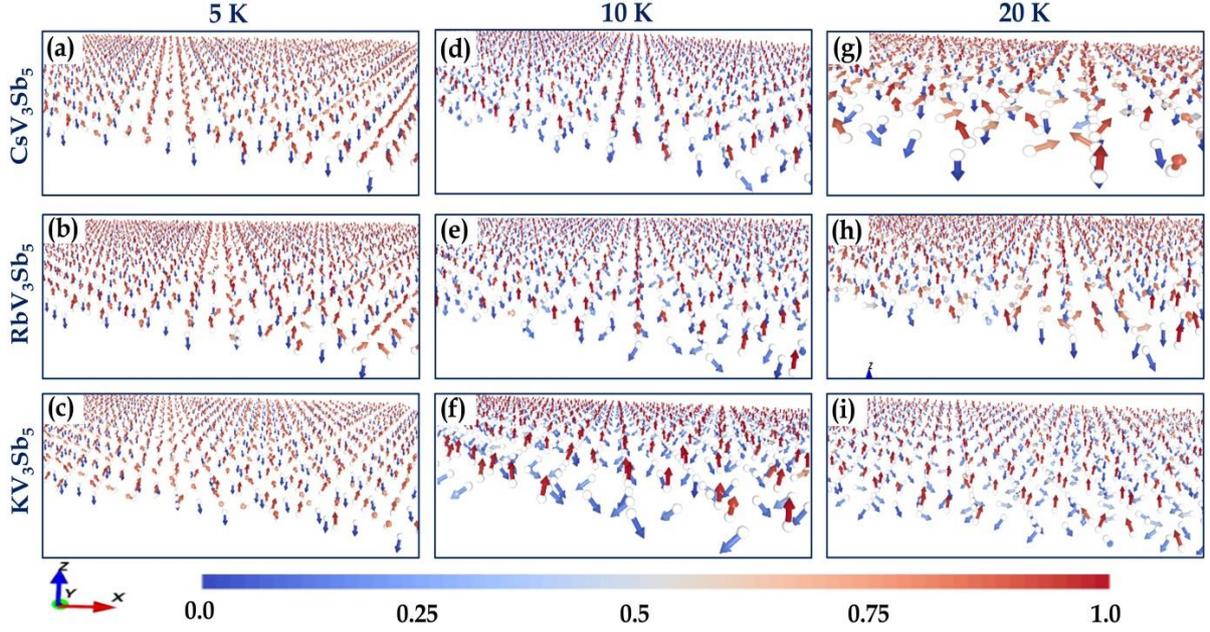

FIG. 5: (a)-(c) The 3D view of the magnetic ground-state spin-textures with at 5 K; (d)-(f) The 3D view of the magnetic ground-state spin-textures at 10 K; (g)-(i) The 3D view of the magnetic ground-state spin-textures at 20 K; for CVS, RVS and KVS respectively. All calculations include Heisenberg exchange and symmetric- and antisymmetric anisotropic exchange interactions as well as magneto crystalline anisotropy.

Presence of such chiral magnetic states at finite temperatures may impart an important impact towards the analysis of the experimental observation of electronic magneto-chiral anisotropic effects in CVS, where, there is a sudden onset of unusually large chiral transport with lowering of temperature below 35K [5]. For such centrosymmetric system, the mirror symmetries are experimentally observed to be inherently broken by the itinerant carriers in the correlated phases. Although, for CVS, the presence of time-reversal symmetry breaking is under scrutiny after the absence of any polar Kerr signal, as observed by Saykin et al. [4], the presence of other spontaneously broken symmetries cannot be ruled out. Occurrence of chiral magnetism and spin-spiral ground states in this series can be a potential reason behind the experimentally observed broken symmetries. It should also be noted that the structural instability and the respective distortion is quite weak and there are disputes and debates accompanied with the yet undetermined low-temperature crystal structure [40-43]. Thus, the presence of such inherent chiral dynamical magnetic structure in the kagome system may provide an answer to the simultaneous and spontaneous occurrence of time-reversal, mirror and $C^6$ rotational symmetry breaking in the CO and SC phases.

In conclusion, for the first part of the comprehensive study of the V-based kagome stibnites [10], we addressed the effects of the formation of the local moments at the V-sites in the kagome series $AV_3Sb_5$ (A = Cs, Rb, K) on their charge-order and superconductivity [10]. In this second part, we have investigated the dynamical magnetic ground state of the system. The underlying arrangement of the spin-magnetic moments reveals that, in addition to the FM long-range order along the [001] direction, there are complex landscapes of spin-spiral configurations in the [010] direction, consisting of commensurate and incommensurate chiral magnetic textures. The presence of such chiral magnetic textures has the capability of spontaneously breaking the time-reversal, mirror or the $C_6$ rotational symmetry, and thus may provide a valid explanation to the experimental findings. In addition, the theoretically derived dynamical structure factors support the presence of complex, collective magnetic excitations within the system, which may be validated by future experimental studies. Thus, our thorough analysis of the dynamical aspects of the magnetism for the V-based kagome stibnites can be helpful in providing an explanation to the existing experimental queries and may also provide a motivation for the future experimental studies.

**Acknowledgements**

Financial support from Vetenskapsrådet (grant numbers VR 2016-05980 and VR 2019-05304), and the Knut and Alice Wallenberg foundation (grant numbers 2018.0060, 2021.0246, 2022.0108 and 2022.0079) is acknowledged. The computations were enabled by resources provided by the National Academic Infrastructure for Supercomputing in Sweden (NAISS) and the Swedish National Infrastructure for Computing (SNIC) at NSC and PDC, partially funded by the Swedish Research Council through grant agreements no. 2022-06725 and no. 2018-05973. OE also acknowledges support from STandUPP and eSSENCE. DK acknowledges BARC supercomputing facility. MNH acknowledges CSIR (India) for fellowship. IDM acknowledges support from the JRG Program at APCTP through the Science and Technology

Promotion Fund and Lottery Fund of the Korean Government, as well as from the Korean Local Governments-Gyeongsangbuk-do Province and Pohang City. IDM and SS also acknowledge financial support from the National Research Foundation of Korea (NRF), funded by the Ministry of Science and ICT (MSIT), through the Mid-Career Grant No. 2020R1A2C101217411.

**References**

[1]     C. Mielke, D. Das, J. X. Yin, H. Liu, R. Gupta, Y. X. Jiang, M. Medarde, X. Wu, H. C. Lei, J. Chang, P. Dai, Q. Si, H. Miao, R. Thomale, T. Neupert, Y. Shi, R. Khasanov, M. Z. Hasan, H. Luetkens, and Z. Guguchia, Nature **602**, 245 (2022).

[2]     Y.-X. Jiang, J.-X. Yin, M. M. Denner, N. Shumiya, B. R. Ortiz, G. Xu, Z. Guguchia, J. He, M. S. Hossain, X. Liu, J. Ruff, L. Kautzsch, S. S. Zhang, G. Chang, I. Belopolski, Q. Zhang, T. A. Cochran, D. Multer, M. Litskevich, Z.-J. Cheng, X. P. Yang, Z. Wang, R. Thomale, T. Neupert, S. D. Wilson, and M. Z. Hasan, Nat. Mater. **20**, 1353 (2021).

[3]     S.-Y. Yang, Y. Wang, B. R. Ortiz, D. Liu, J. Gayles, E. Derunova, R. Gonzalez-Hernandez, L. Šmejkal, Y. Chen, S. S. P. Parkin, S. D. Wilson, E. S. Toberer, T. McQueen, and M. N. Ali, Sci. Adv. **6**, eabb6003 (2020).

[4]     D. R. Saykin, C. Farhang, E. D. Kountz, D. Chen, B. R. Ortiz, C. Shekhar, C. Felser, S. D. Wilson, R. Thomale, J. Xia, and A. Kapitulnik, Phys. Rev. Lett. 131, 016901 (2023).

[5]     C. Guo, C. Putzke, S. Konyzheva, X. Huang, M. Gutierrez-Amigo, I. Errea, D. Chen, M. G. Vergniory, C. Felser, M. H. Fischer, T. Neupert and P. J. W. Moll, Nature 611, 461–466 (2022).

[6]     W. Schweika, M. Valldor, and P. Lemmens, Phys. Rev. Lett. 98, 067201 (2007).

[7]     Daniel Boyko, Avadh Saxena and Jason T Haraldsen, Ann. Phys. (Berlin) 1900350 (2019).

[8]     Mathieu Taillefumier, Julien Robert, Christopher L. Henley, Roderich Moessner, and Benjamin Canals, Phys. Rev. B 90, 064419 (2014).

[9]     Masahide Nishiyama, Satoru Maegawa, Toshiya Inami, and Yoshio Oka Phys. Rev. B 67, 224435 (2003).

[10]    Jonas Becker and Stefan Wessel Phys. Rev. B 100, 241113(R) (2019).

[11]    Tao Xie, Qiangwei Yin, Qi Wang, A. I. Kolesnikov, G. E. Granroth, D. L. Abernathy, Dongliang Gong, Zhiping Yin, Hechang Lei, and A. Podlesnyak, Phys. Rev. B 106, 214436 (2022).

[12]    M. N. Hasan, R. Bharati, J. Hellsvik, A. Delin, S. K. Pal, A. Bergman, S. Sharma, I. D. Marco, M. Pereiro, P. Thunström, P. M. Oppeneer, O. Eriksson, and D. Karmakar, Phys. Rev. Lett.    (accepted),    https://journals.aps.org/prl/accepted/ea077Y18R6f1c990d0859378e012ab2662a4490ae

[13]    M. Ležaić, P. Mavropoulos, G. Bihlmayer, and S. Blügel, Phys. Rev. B **88**, 134403 (2013).

[14]    A. I. Liechtenstein, M. I. Katsnelson, V. P. Antropov, and V. A. Gubanov, J. Magn. Magn. Mater. **67**, 65 (1987).

[15]    Y. O. Kvashnin, O. Grånäs, I. Di Marco, M. I. Katsnelson, A. I. Lichtenstein, and O. Eriksson, Phys. Rev. B **91**, 125133 (2015).

[16]    Y. O. Kvashnin, A. Bergman, A. I. Lichtenstein, and M. I. Katsnelson, Phys. Rev. B **102**, 115162 (2020).

[17]    A. Grechnev, I. Di Marco, M. I. Katsnelson, A. I. Lichtenstein, J. Wills, and O. Eriksson, Phys. Rev. B **76**, 035107 (2007).

[18]    P. Thunström, I. Di Marco, and O. Eriksson, Phys. Rev. Lett. **109**, 186401 (2012).

[19]    J. M. Wills, M. Alouani, P. Andersson, A. Delin, O. Eriksson, and O. Grechnyev, *Full-potential electronic structure method: Energy and force calculations with density functional and dynamical mean field theory* (Springer Science & Business Media, 2010), Vol. 167.

[20]    A. Szilva, Y. Kvashnin, E. A. Stepanov, L. Nordström, O. Eriksson, A. I. Lichtenstein, and M. I. Katsnelson, Rev. Mod. Phys. 95, 035004 (2023).

[21]    M. Udagawa and Y. Motome, Phys. Rev. Lett. 104, 106409 (2010).

[22]    L. Messio, B. Bernu, and C. Lhuillier, Phys. Rev. Lett. 108, 207204 (2012).
[23]    M. Bode, M. Heide, K. von Bergmann, P. Ferriani, S. Heinze, G. Bihlmayer, A. Kubetzka, O. Pietzsch, S. Blügel, and R. Wiesendanger, Nature **447**, 190 (2007).

[24]    B. Zimmermann, M. Heide, G. Bihlmayer, and S. Blügel, Phys. Rev. B **90**, 115427 (2014).

[25]    S. Heinze, K. von Bergmann, M. Menzel, J. Brede, A. Kubetzka, R. Wiesendanger, G. Bihlmayer, and S. Blügel, Nature Phys **7**, 713 (2011).

[26]    N. Romming, C. Hanneken, M. Menzel, J. E. Bickel, B. Wolter, K. von Bergmann, A. Kubetzka, and R. Wiesendanger, Science **341**, 636 (2013).

[27]    A. Neubauer, C. Pfleiderer, B. Binz, A. Rosch, R. Ritz, P. G. Niklowitz, and P. Böni, Phys. Rev. Lett. **102**, 186602 (2009).

[28]    C. Franz, F. Freimuth, A. Bauer, R. Ritz, C. Schnarr, C. Duvinage, T. Adams, S. Blügel, A. Rosch, Y. Mokrousov, and C. Pfleiderer, Phys. Rev. Lett. **112**, 186601 (2014).

[29]    T. Schulz, R. Ritz, A. Bauer, M. Halder, M. Wagner, C. Franz, C. Pfleiderer, K. Everschor, M. Garst, and A. Rosch, Nature Phys **8**, 301 (2012).

[31]    R. M. Otxoa, P. E. Roy, R. Rama-Eiroa, J. Godinho, K. Y. Guslienko, and J. Wunderlich, Commun Phys **3**, 190 (2020).

[32]    K.-S. Ryu, L. Thomas, S.-H. Yang, and S. Parkin, Nature Nanotech. **8**, 527 (2013).

[28]    S. Emori, U. Bauer, S.-M. Ahn, E. Martinez, and G. S. D. Beach, Nature Mater **12**, 611 (2013).

[33]   C. Etz, L. Bergqvist, A. Bergman, A. Taroni, and O. Eriksson, J. Phys.: Condens. Matter **27**, 243202 (2015).

[34]   B. Skubic, J. Hellsvik, L. Nordström, and O. Eriksson, J. Phys.: Condens. Matter **20**, 315203 (2008).

[35]   P. Kurz, F. Förster, L. Nordström, G. Bihlmayer, and S. Blügel, Phys. Rev. B **69**, 024415 (2004).

[36]   J. C. Leiner, T. Kim, K. Park, J. Oh, T. G. Perring, H. C. Walker, X. Xu, Y. Wang, S. W. Cheong, and J.-G. Park, Phys. Rev. B **98**, 134412 (2018).

[37]   S. Petit, JDN **12**, 105 (2011).

[38]   S. Toth and B. Lake, J. Phys.: Condens. Matter **27**, 166002 (2015).

[39]   See Supplemental materials for the description of the commensurate and incommensurate spirals, the $120^0$ structures, the spin-textures with different terms of the full spin-Hamiltonian and the corresponding dynamical structure factor plots and the temperature dependent spin-textures.

[40]   H. Zhao, H. Li, B. R. Ortiz, S. M. L. Teicher, T. Park, M. Ye, Z. Wang, L. Balents, S. D. Wilson and I. Zeljkovic, Nature 599, 216–221 (2021).

[41]   B. R. Ortiz, S. M. L. Teicher, Y. Hu, J. L. Zuo, P. M. Sarte, E. C. Schueller, A. M. M. Abeykoon, M. J. Krogstad, S. Rosenkranz, R. Osborn, R. Seshadri, L. Balents, J. He, and S. D. Wilson, Phys. Rev. Lett. 125, 247002 (2020).

[42]   J. Luo, Z. Zhao, Y. Z. Zhou, J. Yang, A. F. Fang, H. T. Yang, H. J. Gao, R. Zhou and G-Q Zheng, npj Quantum Mater. 7, 30 (2022).

[43]   Q. Stahl, D. Chen, T. Ritschel, C. Shekhar, E. Sadrollahi, M. C. Rahn, O. Ivashko, M. v. Zimmermann, C. Felser, and J. Geck, Phys. Rev. B 105, 195136 (2022).

Supplementary Materials for

# Magnetism in $AV_3Sb_5$ (A = Cs, Rb, K): Complex Landscape of the Dynamical Magnetic Textures

Debjani Karmakar[1,2,#,*], Manuel Pereiro[1,#], Md. Nur Hasan[3], Ritadip Bharati[4], Johan Hellsvik[5], Anna Delin[6,7], Samir Kumar Pal[3], Anders Bergman[1], Shivalika Sharma[8], Igor Di Marco[1,8,9], Patrik Thunström[1], Peter M. Oppeneer[1] and Olle Eriksson[1,*]

[1] *Department of Physics and Astronomy,*
*Uppsala University, Box 516, SE-751 20, Uppsala, Sweden*

[2] *Technical Physics Division*
*Bhabha Atomic Research Centre, Mumbai 400085, India*

[3] *Department of Chemical and Biological Sciences,*
*S. N. Bose National Centre for Basic Sciences,*
*Block JD, Sector-III, SaltLake, Kolkata 700 106, India*

[4] *School of Physical Sciences*
*National Institute of Science Education and Research*
*HBNI, Jatni - 752050, Odisha, India*

[5] *PDC Center for High Performance Computing,*
*KTH Royal Institute of Technology, SE-100 44 Stockholm, Sweden*

[6] *Department of Applied Physics*
*KTH Royal Institute of Technology, SE-106 91 Stockholm, Sweden*

[7] *Swedish e-Science Research Center (SeRC),*
*KTH Royal Institute of Technology, SE-10044 Stockholm, Sweden*

[8] *Asia Pacific Center for Theoretical Physics,*
*Pohang, 37673, Republic of Korea*

[9] *Department of Physics, Pohang University of Science and Technology,*
*Pohang, 37673, Republic of Korea*

[#] D.K. and M.P. contributed equally

[*]Corresponding Authors:
   Debjani Karmakar
   Email: debjani.karmakar@physics.uu.se

   Olle Eriksson
   Email: olle.eriksson@physics.uu.se

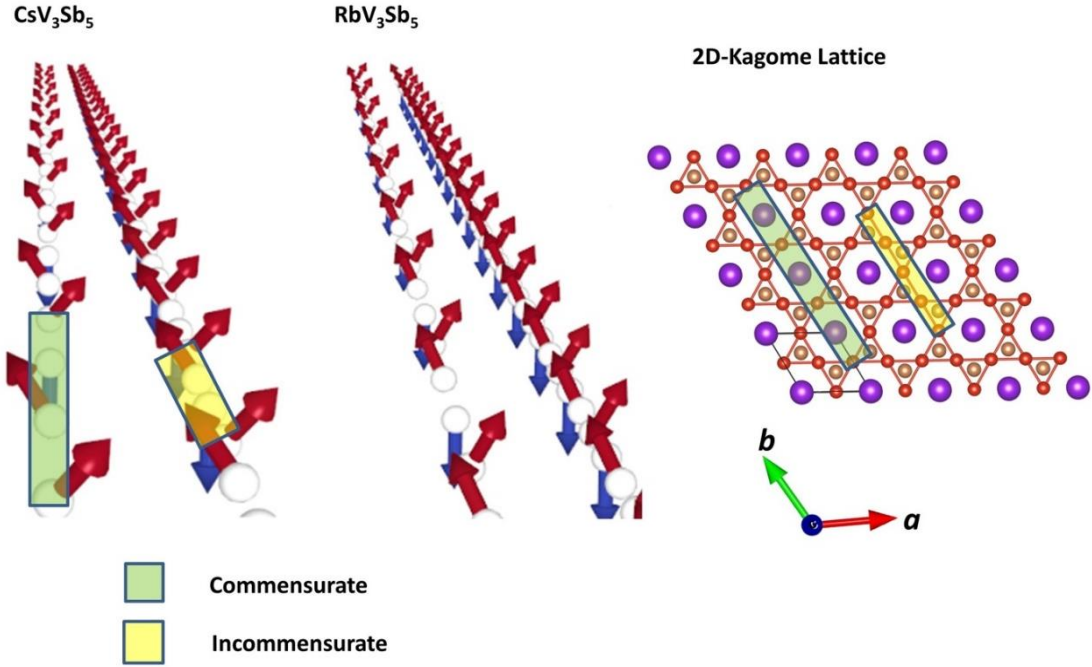

FIG. S1: The commensurate and incommensurate spin-spirals for $CsV_3Sb_5$ and $RbV_3Sb_5$. The highlighted green and yellow boxes correspond to the pitches of the commensurate and incommensurate spirals respectively. On the right hand panel, we have shown the corresponding pitches on the 2D kagome lattice plane. The parallelogram at the left corner of the 2D-lattice shows the unit cell having the dimension of the lattice parameter *a*. The pitches of the commensurate and incommensurate spirals are 3*a* and 1.5*a* respectively, as can be seen from the figure.

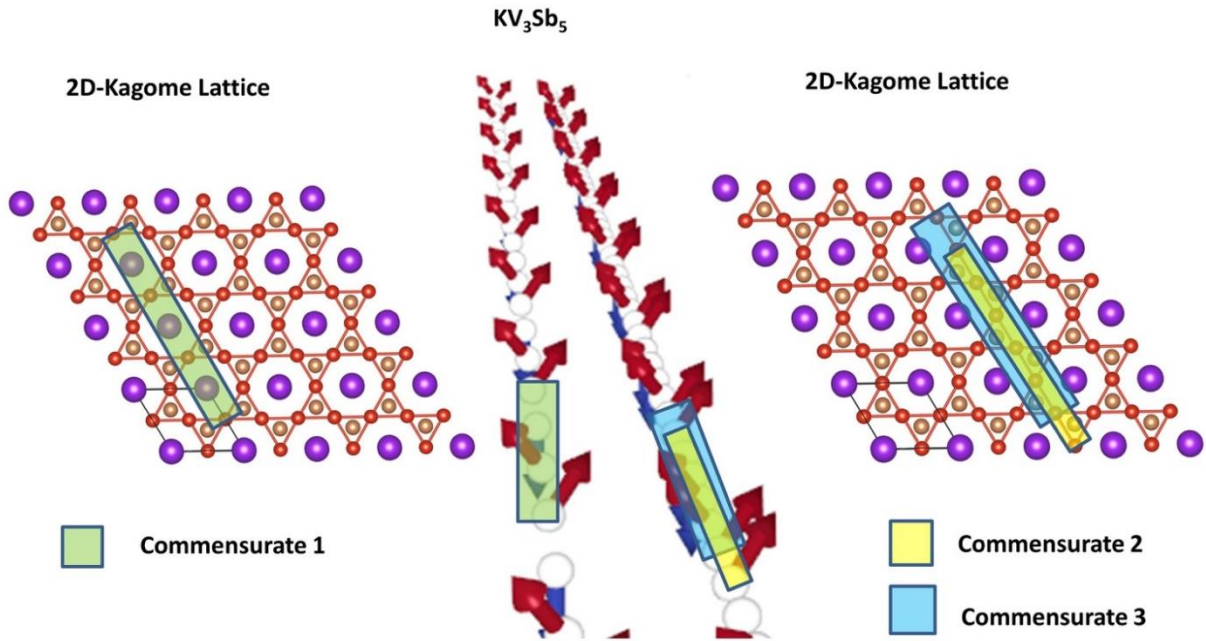

FIG. S2: The middle panel shows the 3D view of the three commensurate spin-spirals for $KV_3Sb_5$. The highlighted green, yellow and cyan boxes correspond to the pitches of the three commensurate spirals respectively. On the left and right hand panel, we have shown the corresponding pitches on the 2D kagome lattice planes. The parallelogram at the left corner of the 2D-lattice shows the unit cell having the dimension of the lattice parameter $a$. The pitches of all the three commensurate spirals are $3a$, as can be seen from the figure.

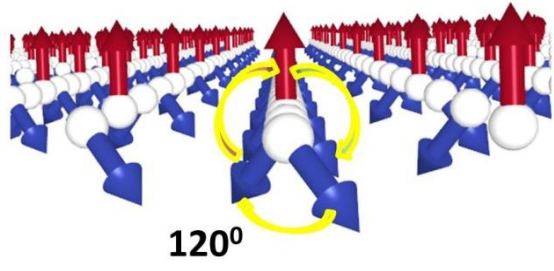

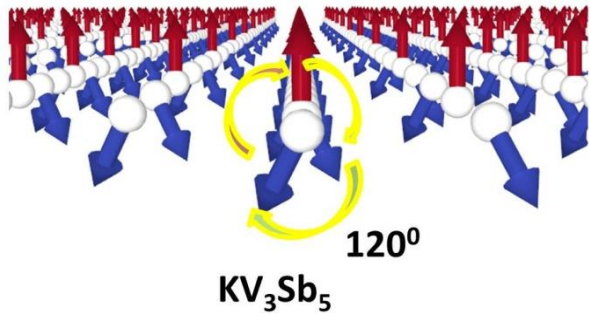

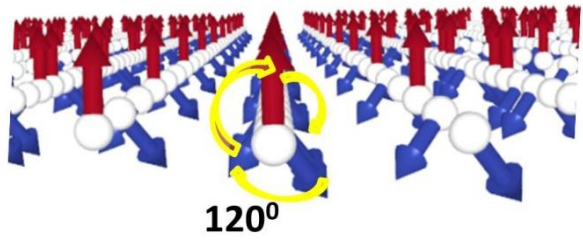

FIG. S3: The three-dimensional (3D)-$120^0$ magnetic textures for CVS, RVS and KVS.

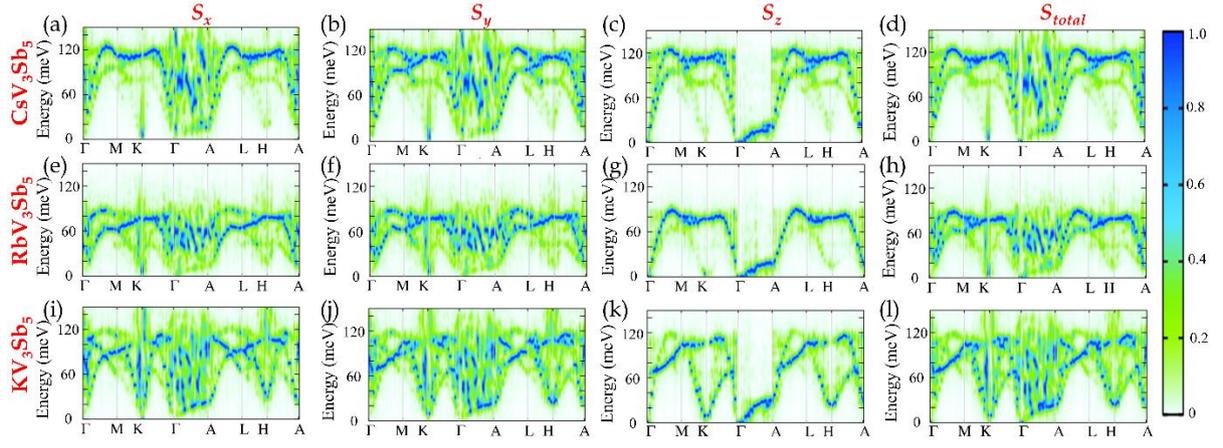

FIG. S4: The spin-component projected and total dynamical structure factors $S(q,\omega)$ with Heisenberg exchange of (a-d) CsV$_3$Sb$_5$, (e-h) RbV$_3$Sb$_5$ and (i-l) KV$_3$Sb$_5$.

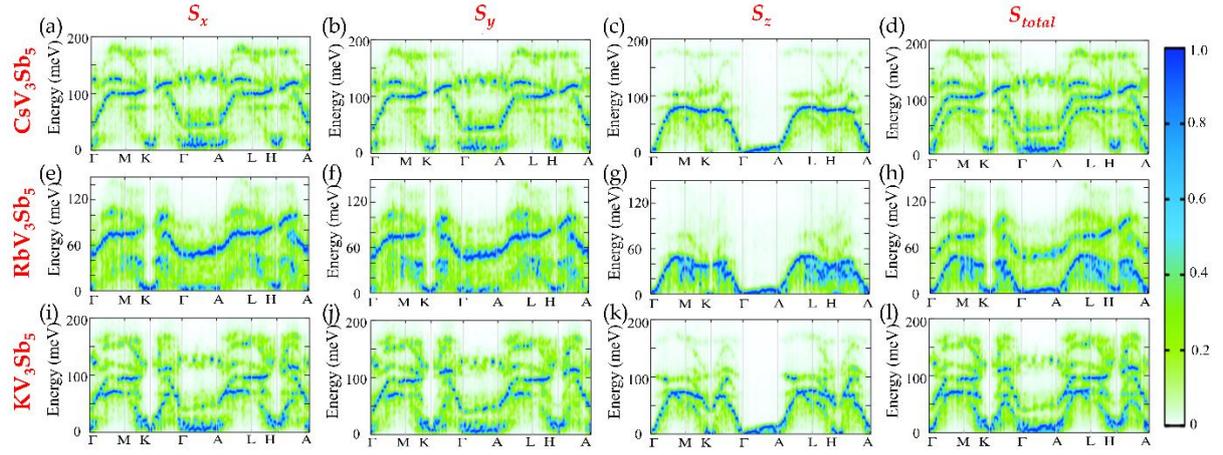

FIG. S5: The spin-component projected and total dynamical structure factors $S(q,\omega)$ with Heisenberg exchange and magneto crystalline anisotropy of (a-d) $CsV_3Sb_5$, (e-h) $RbV_3Sb_5$ and (i-l) $KV_3Sb_5$.

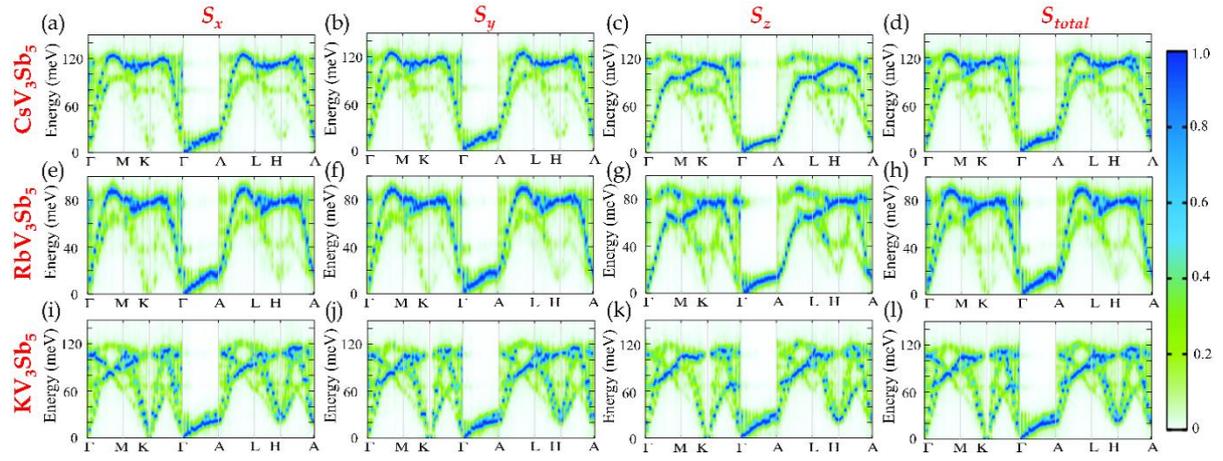

FIG. S6: The spin-component projected and total dynamical structure factors $S(q,\omega)$ with Heisenberg exchange and symmetric anisotropic exchange interactions of (a-d) $CsV_3Sb_5$, (e-h) $RbV_3Sb_5$ and (i-l) $KV_3Sb_5$.

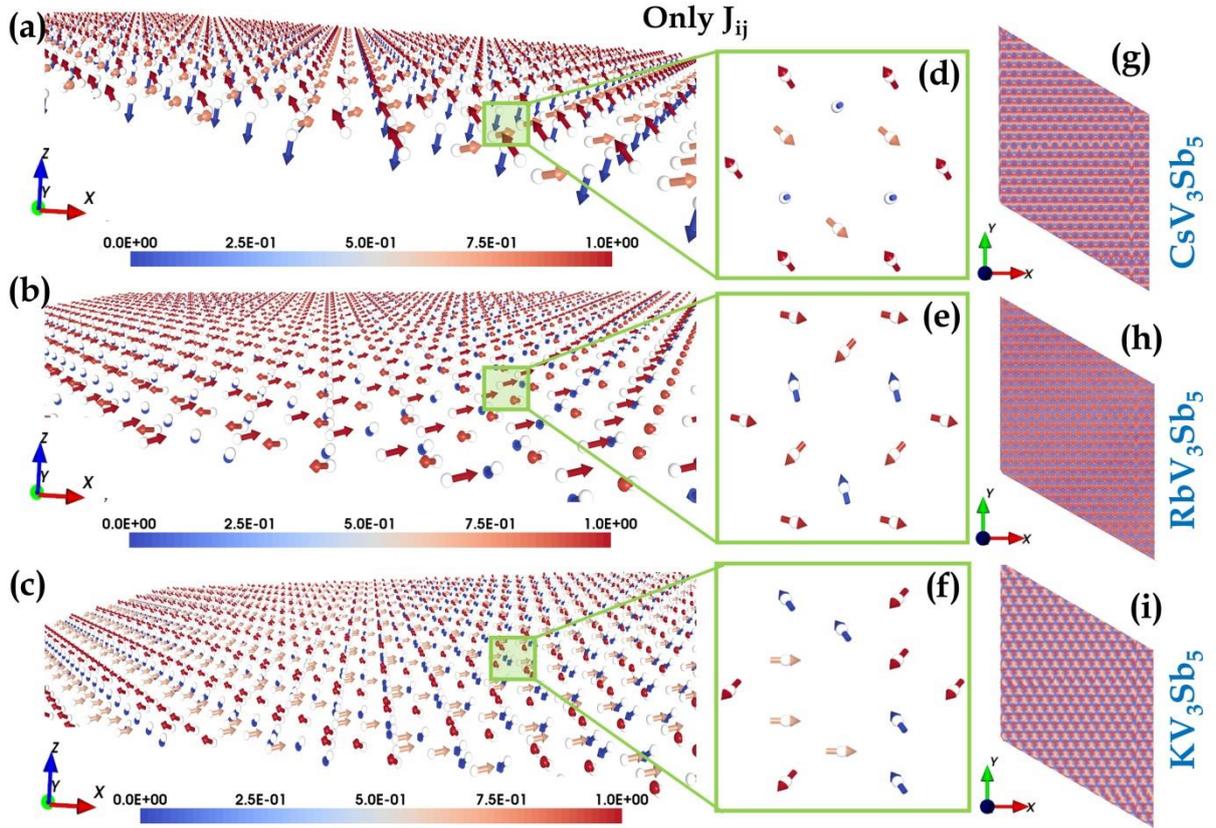

FIG. S7: (a)-(c) The 3D view of the dynamical ground-state spin-textures with only Heisenberg exchange; (d)-(f) spin-configurations on the star-of-David of the kagome lattice; (g)-(i) the corresponding magnetization densities for CVS, RVS and KVS respectively.

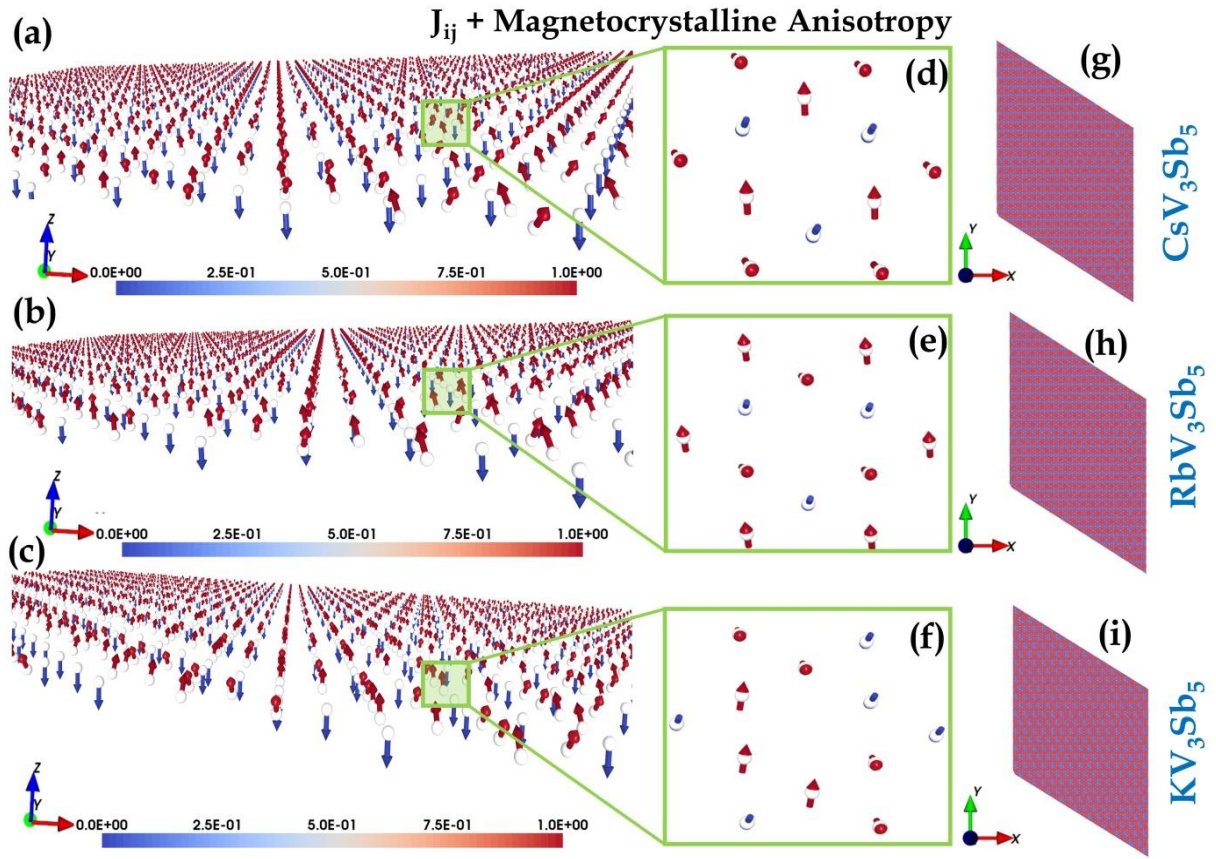

FIG. S8: (a)-(c) The 3D view of the dynamical ground-state spin-textures with Heisenberg exchange and magneto crystalline anisotropy; (d)-(f) spin-configurations on the star-of-David of the kagome lattice; (g)-(i) the corresponding magnetization densities for CVS, RVS and KVS respectively.

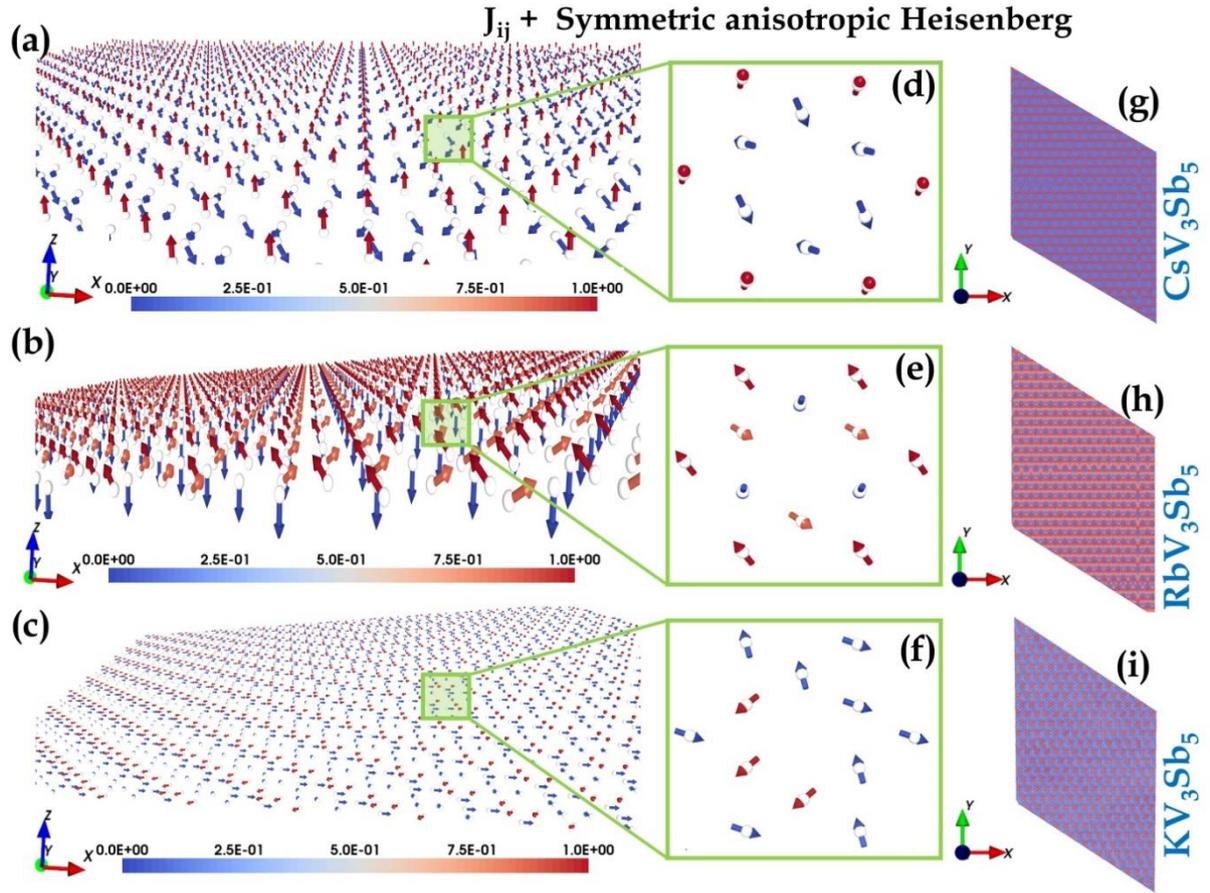

FIG. S9: (a)-(c) The 3D view of the dynamical ground-state spin-textures with Heisenberg exchange and symmetric anisotropic exchange interactions; (d)-(f) spin-configurations on the star-of-David of the kagome lattice; (g)-(i) the corresponding magnetization densities for CVS, RVS and KVS respectively.